\documentclass[aps,pr
,preprint
,showpacs
,groupedaddress
,superscriptaddress
,nobibnotes
]
{revtex4-1}
\usepackage{amsmath,amsthm,amssymb}
\usepackage{array}
\usepackage{subfigure}
\usepackage{graphicx}
\usepackage{fancyhdr}
\usepackage{color}
\usepackage{extarrows}
\usepackage{enumerate}
\usepackage{epstopdf}
\usepackage{bm}
\usepackage[colorlinks,
          linkcolor=black,
            citecolor=black,
            urlcolor=blue
           ]{hyperref}
\usepackage{natbib}
\usepackage{comment}

\usepackage{xr}
\usepackage{tabularx}
\begin{document}
\title{Hall anomaly by vacancies in pinned lattice of vortices: A quantitative analysis on the thin-film data of BSCCO}
\author{Ruonan Guo}
\affiliation{Shanghai Center for Quantitative Life Sciences \& Physics Department, Shanghai University, Shanghai 200444, China}
\affiliation{Shanghai Key Laboratory of High Temperature Superconductors, Shanghai University, Shanghai 200444, China}
\author{Yong-Cong Chen}
\email{chenyongcong@shu.edu.cn}
\affiliation{Shanghai Center for Quantitative Life Sciences \& Physics Department, Shanghai University, Shanghai 200444, China}
\author{Ping Ao}
\email{aoping@sjtu.edu.cn}
\affiliation{Shanghai Center for Quantitative Life Sciences \& Physics Department, Shanghai University, Shanghai 200444, China}
\begin{abstract}
Hall anomaly, as appears in the mixed-state Hall resistivity of type-II superconductors, has had numerous theories but yet a consensus on its origin. In this work, we conducted a quantitative analysis of the magnetotransport measurements on BSCCO thin films by Zhao et al. [Phys. Rev. Lett. 122, 247001 (2019)] and validate a previously proposed vacancy mechanism [cf. J. Phys. Condens. Matter. 10, L677 (1998)] with many-body vortex correlations for the phenomenon. The model attributes the Hall anomaly to the motion of vacancies in pinned fragments of vortex lattice. Its validity is first examined by an exploration on the vortex states near the Kosterlitz-Thouless transition on the vortex crystal. Comparisons are then carried out between the measured activation energies with the calculated creation energy of the vortex-anti-vortex pair and the vacancy energy on the flux-line lattice, with no adjustable parameter. Our analysis elucidates the theoretical basis and prerequisites of the vacancy model. In particular, the vacancy activation energies are an order of magnitude smaller than that of a sole vortex line. The proposed mechanism may provide a macro-theoretical framework for other studies.

\end{abstract}

\maketitle

\renewcommand{\vec}[1]{\mathbf{#1}}
\newcommand{\bvec}[1]{\mbox{\boldmath $#1$}}
\newcommand{\cmmnt}[1]{\ignorespaces}

\section{Introduction}

The Hall anomaly in the mixed-state Hall resistivity of superconductors, i.e. the sign reversal of the Hall resistivity below the superconducting transition temperature and in the presence of flux-line vortices, was discovered as early as in the 1950s \cite{niessen1965hall,reed1965observation}. Much attention has since been set on its physical origin. Prior to the discovery, Onsager \cite{onsager1949motion} had established the framework of the vortex theory based on fluid dynamics in 1949. A modern version of his work, the equation for a $j^{th}$ vortex of unit length in a superconductor takes the same Langevin equation as a charged particle in the presence of a magnetic field \cite{ao1998motion},
\begin{equation}\label{Langevin}
 m\ddot{\vec{r}}_j = q\left(\frac{n_{s}}{2}\right)h\left(\vec{v}_{s,t}-\dot{\vec{r}}_j\right)\times \hat{\vec{z}}-\eta \dot{\vec{r}}_j+\vec{F}_p+\vec{f},
\end{equation}
where the overhead of dots stands for time derivatives. The unit length vortex at $\vec{r_j}$ has an effective mass $m$, subject to a pinning force $\vec{F}_p$, a fluctuating force $\vec{f}$, viscosity $\eta$, and moves in a background of a superfluid with total velocity $\vec{v}_{s,t}$ (which includes contributions from all other vortices). For the parameters in Eq.~(\ref{Langevin}), $q = \pm 1$ indicates the vorticity (under the usual right-hand rules), $h$ is the Planck constant, $n_s$ the superfluid carrier density, and $\vec{\hat{z}}$ the unit vector in the direction of the magnetic field. The term with the velocity $\dot{\vec{r}}_j$ at the right-hand side is also known as the Magnus force. Note that CGS units are assumed throughout this work, in line with the majority of work in the literature.

\begin{figure}[!ht]
\begin{tabular}{cc}
{\label{Fig3(a)}
\includegraphics[width=0.5\textwidth]{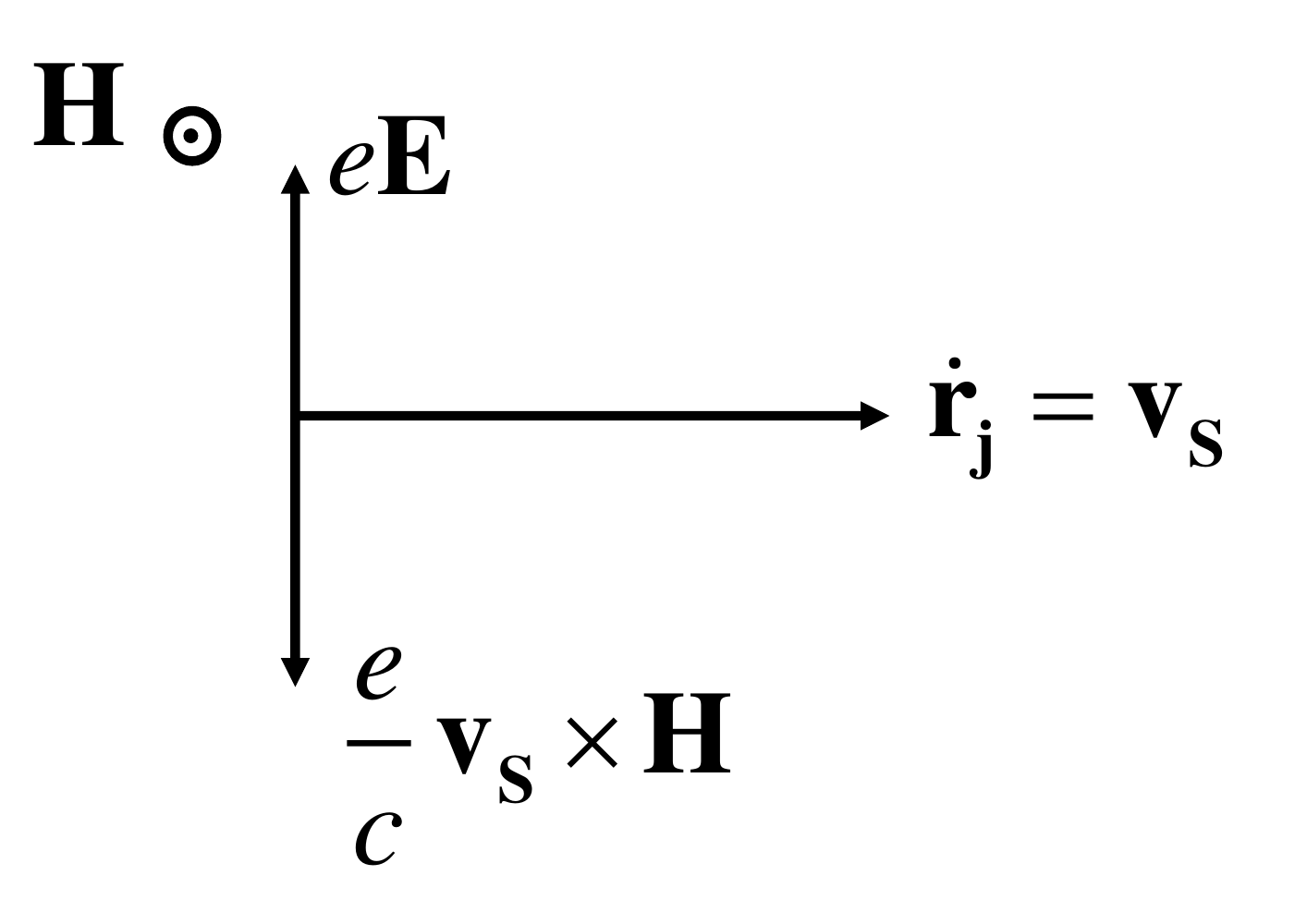}}&
{\label{Fig3(b)}
\includegraphics[width=0.5\textwidth]{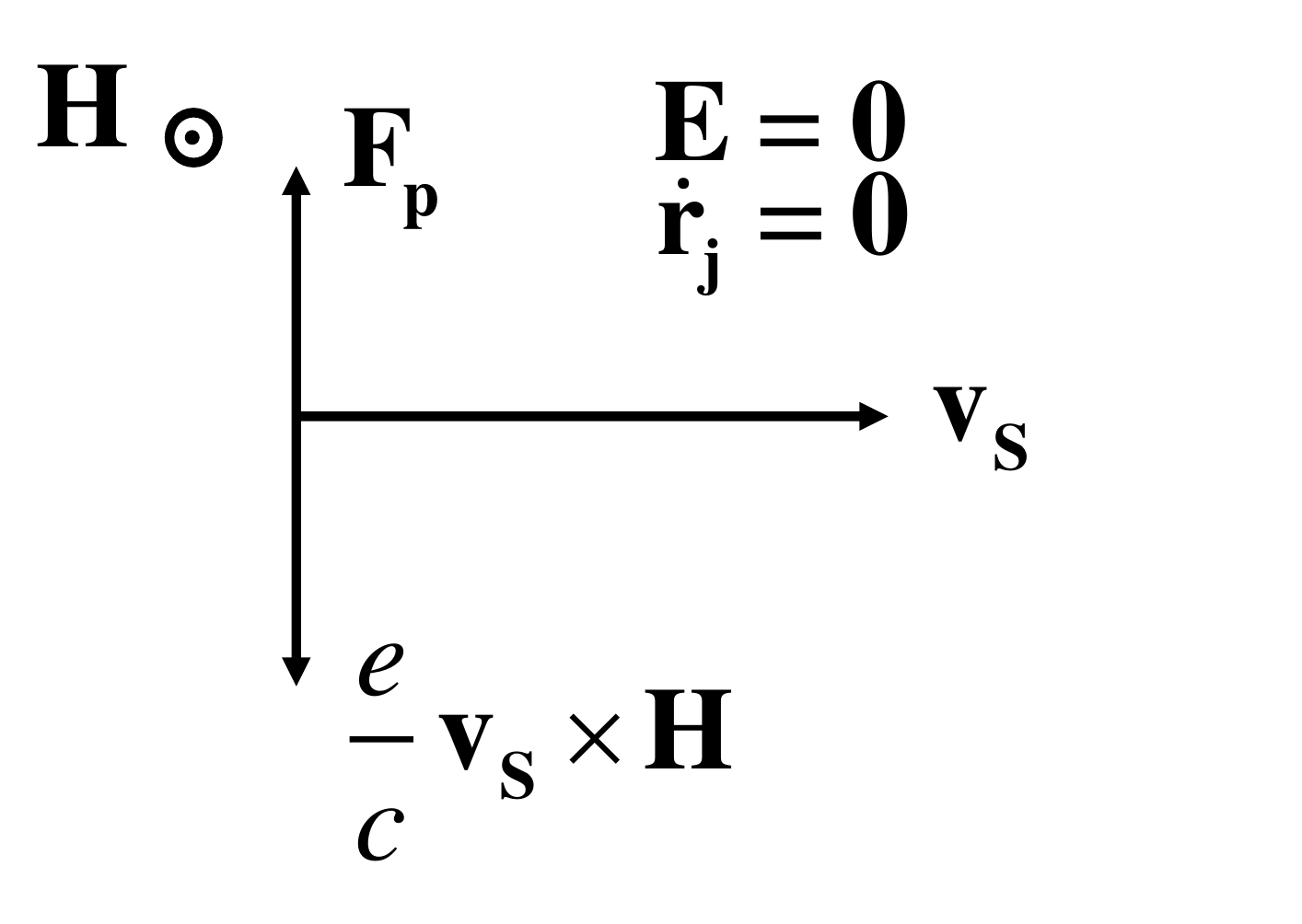}}\\
(a) & (b)
\end{tabular}
\protect\caption{Analysis of vortex dynamics in two ideal situations. In the picture $\vec{E}$ is the electric field generated by moving vortex and $\vec{H}$ is the applied magnetic field. (a) With no pinning force $\vec{F}_p=0$ and no frictional force $(\eta = 0; \vec{f}=0)$, the vortex moves at the same velocity as the superfluid $\dot{\vec{r}}_j = \vec{v}_s$. (b) The vortex is firmly trapped by strong pinning force $\vec{F}_p$ such that $\dot{\vec{r}}_j=0$.
}
\label{Fig3}
\end{figure}

With Eq.~(\ref{Langevin}), two idealized pictures can be drawn. Take e.g. $q = +1$, Fig.~\ref{Fig3}(a) shows that in the absence of the pinning force $\vec{F}_p$ and the frictional force $(\eta = 0; \vec{F}_p = 0)$, the vortex velocity $\dot{\vec{r}}_j$ matches that of superfluid in both direction and magnitude. In other words, for an average charge $e$ in the vortex, the Magnus Force (due to the electric field $\vec{E}$ generated by moving vortex), i.e. $e\vec{E}$ cancels the Lorentz force $(e/c)\vec{v}_s\times \vec{H}$, leading to the same Hall effect as in a normal metal. One has no reason to expect a sign reversal on the Hall resistance $R_{xy}$ below the superconducting transition temperature $T_c$. Fig.~\ref{Fig3}(b) presents an alternative scenario of an extreme situation: The vortex is firmly trapped by a strong pinning force $\vec{F}_p$ such that there is no Magnus force. Moreover, the pinning force is opposite to the Lorentz force and there will be no change of sign on the Hall resistance $R_{xy}$ either. These scenarios raise an apparent paradox between conceptual reasoning and experimental observation.

To address the abnormal sign reversal, several groups \cite{nozieres1966motion,hagen1990anomalous,hagen1993anomalous,wang1991anomalous} have proposed a phenomenological model, which adds an adjustable parameter $\alpha$ directly in front of the Magnus force. This modifies the vortex dynamic equation to
\begin{equation}\label{Langevin,2}
 m\ddot{\vec{r}}_j = q\left(\frac{n_s}{2}\right)h(\vec{v}_{s,t}-\alpha \dot{\vec{r}}_j)\times \hat{\vec{z}}-\eta \dot{\vec{r}}_j+\vec{F}_p+\vec{f}.
\end{equation}
In a steady-state where the vortex is subject to the friction force only, the Hall resistivity $\rho_{xy}(0)$ and the longitudinal resistivity $\rho_{xx}(0)$ can be obtained respectively
\begin{eqnarray}
\label{resistivity0,1}
 \rho_{xx}(0) &=& -\frac{\eta qhH}{ec\left(\alpha^2 q^2{n_s}^2h^2 + 4\eta^2\right)},\\
\label{resistivity0,2}
 \rho_{xy}(0) &=& \frac{\alpha n_sq^2h^2H}{2ec\left(\alpha^2 q^2{n_s}^2h^2 + 4\eta^2\right)}.
\end{eqnarray}
This allows the sign reversal of the Hall resistivity $\rho_{xy}(0)$ when $\alpha < 0$, though it is unclear how $\alpha$ can be so adjusted. Furthermore, there should also be resistivity coming from the normal component of electrons at finite temperatures. The latter follows a parallel connection to the superconducting circuit so that we have
\begin{eqnarray}
\label{resistivity,1}
\label{resistivity,1}
  \rho_{xx}(T) &=& -\frac{\eta\eta'qhH}{e\left[c\eta'\alpha^2q^2n_s^2(T)h^2-eh\eta qH(n_s(0)-n_s(T))+4c\eta'\eta^2\right]},\\
\label{resistivity,2}
   \rho_{xy}(T) &=& \frac{\alpha Hn_s(T)q^2h^2}{ce\left[2\alpha^2q^2n_s^2(T)h^2+\alpha n_s(T)q^2h^2(n_s(0)-n_s(T))+8\eta^2\right]}.
\end{eqnarray}
Here $n_s(T)$ is the density of superfluid electron at temperature $T$ and the density of normal electron can be written as $n_s(0)-n_s(T)$. The change in the sign of $\alpha$ has to overcome the normal-state electron contribution to achieve the sign reversal on $\rho_{xy}(T)$. The details are presented in Part A of the Supplemental Material (SM) to this work\cite{supplementary}.

The adjustable parameter $\alpha$ itself has not had a consistent origin in the approach. Hall and Vinen \cite{hall1956rotationII,hall1956rotation} in 1956 assumed a big drag force of the same as the Magnus force. Nozieres and Vinen \cite{nozieres1966motion}, Hagen et al. \cite{hagen1990anomalous,hagen1993anomalous} further developed a theoretical explanation for the existence of such a force. In contrast, Wang and Ting \cite{wang1991anomalous} argued that the parameter $\alpha$ should be zero. They based their theory on the well-known normal-core model for flux lines in extreme type-II superconductors and correctly took into account the backflow current due to pinning forces, which concluded in no Magnus force. As a result, there can be a small $\alpha$ with either sign, leading to a signing inversion of the Hall coefficient under certain circumstances. However, Ao and Thouless \cite{ao1993berry} in 1993 found that $\alpha \equiv 1$ at zero temperature. And three years later, Thouless, Ao and Niu \cite{thouless1996transverse} extended the result to low but finite temperature. Additionally, from the Ginzburg-Landau theory, another phenomenological model was proposed by Xu et al. \cite{xu2002scaling} in 2002 for the Hall anomaly in High-temperature superconductors. And in 2004, Ghenim et al. \cite{shi4} also suggested a model for conventional superconductors.

Despite that the Hall anomaly attracted numerous theoretical studies, a lack of advancement in experimental techniques has hindered a consensus on its origin. But recently, there has been a steady improvement in the situation over the last two decades. In 1996, Zhu et al. \cite{zhu1997observation} designed a mechanical experiment to directly measure the total transverse force on moving vortices in a type II superconductor for the first time, and their result is consistent with the Ao-Thouless theory \cite{ao1993berry}. In 2001, Zhu and Nyeanchi \cite{zhu2001correlation} demonstrated that certain features of the Kosterlitz-Thouless transition of a vortex lattice are preserved near the superconducting transition temperature. In 2019, Chen's team \cite{yu2019high} developed a fabrication process which can produce intrinsic monolayer crystals of BSCCO. In 2021, Richter et al. \cite{richter2021resistivity} measured the resistivity, Hall effect, and anisotropic superconducting coherence lengths in HBCCO thin films with morphological variations. Excellent measurement of Hall effects in an atomically thin high-temperature superconductor by a Harvard group \cite{zhao2019sign} in 2019 drastically extended the region which displays the Hall sign reversal.

In this work, we present an in-depth analysis of the activation energies deduced from the experimental data in \cite{zhao2019sign}. It is then compared to the predicted activation energy of independent vortices and the energy of vortex many-body correlation under flux-line lattices. In the next section, we examine whether the abnormal Hall effect on the BSCCO film meets the pre-requisites of the vacancy model by analyzing the state of the vortices near the KT transition temperature. In section III, we first review the fundamentals of our methodology, namely the pinning and dynamics of vacancies in a vortex lattice. The core concept presented follows throughout the entire section. The experimental data are extracted and compared with the theoretical calculations of activation energies of carriers in the BSCCO film, under varying magnetic fields. The excellent agreement between them elucidates the conformation of the Hall anomaly to the vacancy model. Some concluding remarks and possible connections to other works and future directions are discussed in the final section.

\section{Vortex states around the Kosterlitz-Thouless transition}

To validate that the Hall anomaly complies with the pre-requisites of the vacancy model, it is crucial to clarify the states of vortex crystal near the Kosterlitz-Thouless (KT) transition.  In this section, we will first review some basics of the vortex lattice state near the KT transition which melts the crystal. Then we discuss the nature of the KT transition from the thermodynamic perspective and estimate the theoretical value of the transition temperature $T_{KT}$ of the BSCCO thin film. Finally, the question of whether the concept and presence of vacancies still apply above $T_{KT}$ is addressed.

\subsection{KT transition of molten crystals}

The melting transition of most solid materials at the present has not been well understood as there is a lack of theories explaining the transition on the microscopic scale. Furthermore, the mechanism of melting depends on the interaction details between the constituents forming a crystal lattice. In particular various defects which reduce the translational order of the crystal play a major role. It should be noted that in two dimensions, only edge dislocations and not the screw ones are important in the melting transition. The core energy of the dislocations upon which these effects can form must be sufficiently low for their spontaneous appearance. The specific analysis of the dislocation formation energy in a 2D BSCCO film will be presented in the subsequent section.

These dislocations in question are topological point defects, which implies that a single one cannot be created isolated by an affine transformation without cutting the hexagonal crystal up to infinity (i.e. up to its borders). Hence they must be created in pairs with antiparallel Burgers vectors. When a large number of dislocations were e.g. thermally excited, the discrete translational order of the crystal would be destroyed. Simultaneously, the shear modulus and Young's modulus would disappear, signaling the starting of the molten transition from a solid to a fluid phase.

However, it is possible that the orientational order is not yet fully destroyed (as indicated by lattice lines in one direction) and one finds - very similar to liquid crystals - a fluid phase with typically a six-folded direction field. This so-called hexatic phase still has an orientational stiffness. Such an anisotropic fluid phase can appear when the dislocations dissociate into isolated five-folded and seven-folded disclinations \cite{gasser2010melting}. This two-step melting phenomenon is described within the so-called Kosterlitz-Thouless-Halperin-Nelson-Young-theory (KTHNY theory), based on two separate transitions of the Kosterlitz-Thouless-type. In 2010, Urs Gasser et al. \cite{gasser2010melting} presented the first conclusive evidence for the existence of the hexatic phase and two continuous phase transitions in 2D melts in a colloidal model system with repulsive magnetic dipole-dipole interaction.

\subsection{Thermodynamics of KT transition}
The topological phase transition in a 2D superfluid was predicted by Berezinskii \cite{berezinskii1970destroying} and Kosterlitz and Thouless \cite{kosterlitz1973ordering} and elaborated by Halperin and Nelson \cite{nelson1979dislocation,nelson1978study}. A simple thermodynamic argument allows us to understand the intrinsic quality of the KT transition \cite{blatter1994vortices}. The Helmholtz free energy is given by the difference between the energy $E$ and the entropy of a dislocation $S$ multiplied by the temperature $T$,
\begin{equation}\label{free energy}
 F = E-TS.
\end{equation}
The energy $E$ is given by Eq.~(\ref{a dislocation pair}), which is contributed by a dislocation pair over a large distance. For convenience, we rewrite it as
\begin{equation}\label{free energy}
 E = \frac{d\varepsilon_0}{2\sqrt{3}\pi}\ln\left(\frac{L}{a}\right).
\end{equation}
The entropy can be estimated from the number of places the dislocation can be positioned, namely on each of the $\sim L^2$ plaquette of the lattice, i.e.,
\begin{equation}\label{entropy}
 S = k_{B}\ln\left(\frac{L^2}{a^2}\right).
\end{equation}
Accordingly, the free energy is given by
\begin{equation}\label{free energy re}
 F = \left[d\frac{\varepsilon_0}{2\sqrt{3}\pi}-2k_{B}T\right] \ln\left(\frac{L}{a}\right).
\end{equation}
Evidently, there exists a temperature above which a vast number of dislocations are preferred. Such a transition is of topological nature and is referred to as the KT transition. The transition temperature $T_{KT}$ could be expressed as
\begin{equation}\label{theory}
 T_{KT} = \frac{1}{4\sqrt{3}\pi k_{B}}\left(\frac{\Phi_0}{4\pi \lambda}\right)^2d \cong 57.3 K.
\end{equation}

Ideally at $T<T_{KT}$, thermally excited dislocations form pairs of close compact, namely a dislocation pair.  But they are spontaneously separated when $T>T_{KT}$. The KT transition temperature $T_{KT}$ measured by a Harvard group is 60 K \cite{zhao2019sign}, which validates the effective thickness and the London penetration depth chosen below for our analysis.

\subsection{The presence of vacancies above $T_{KT}$}

Let us for the moment assume a vortex solid-liquid phase transition near (the first) $T_{KT}$, around 60 K. Then the Arrhenius behavior of longitudinal resistance places the Hall sign reversal within the thermally activated flux flow regime above the vortex lattice melting temperature \cite{ao2020comment}. An urgent question is whether the Hall anomaly mainly appears in the vortex-liquid regime. Scilicet fades away as the vortex liquid freezes into a solid-state crystal.

However, there is no evidence for such a solid-liquid phase transition in the BSCCO film. As the transition would be a first-order phase transition in which observation of latent heat was to be expected, the resistance of the BSCCO film should measure an abrupt change at the transition. More crucially Kosterlitz-Thouless transition is classified as a topological phase transition, the third type of phase transition. Experientially, all curves intercepted by the isotherm of $T_{KT} = 60$ K are smooth, with no sign of jump around the node.

\begin{figure}
{
\includegraphics[width=0.8\textwidth]{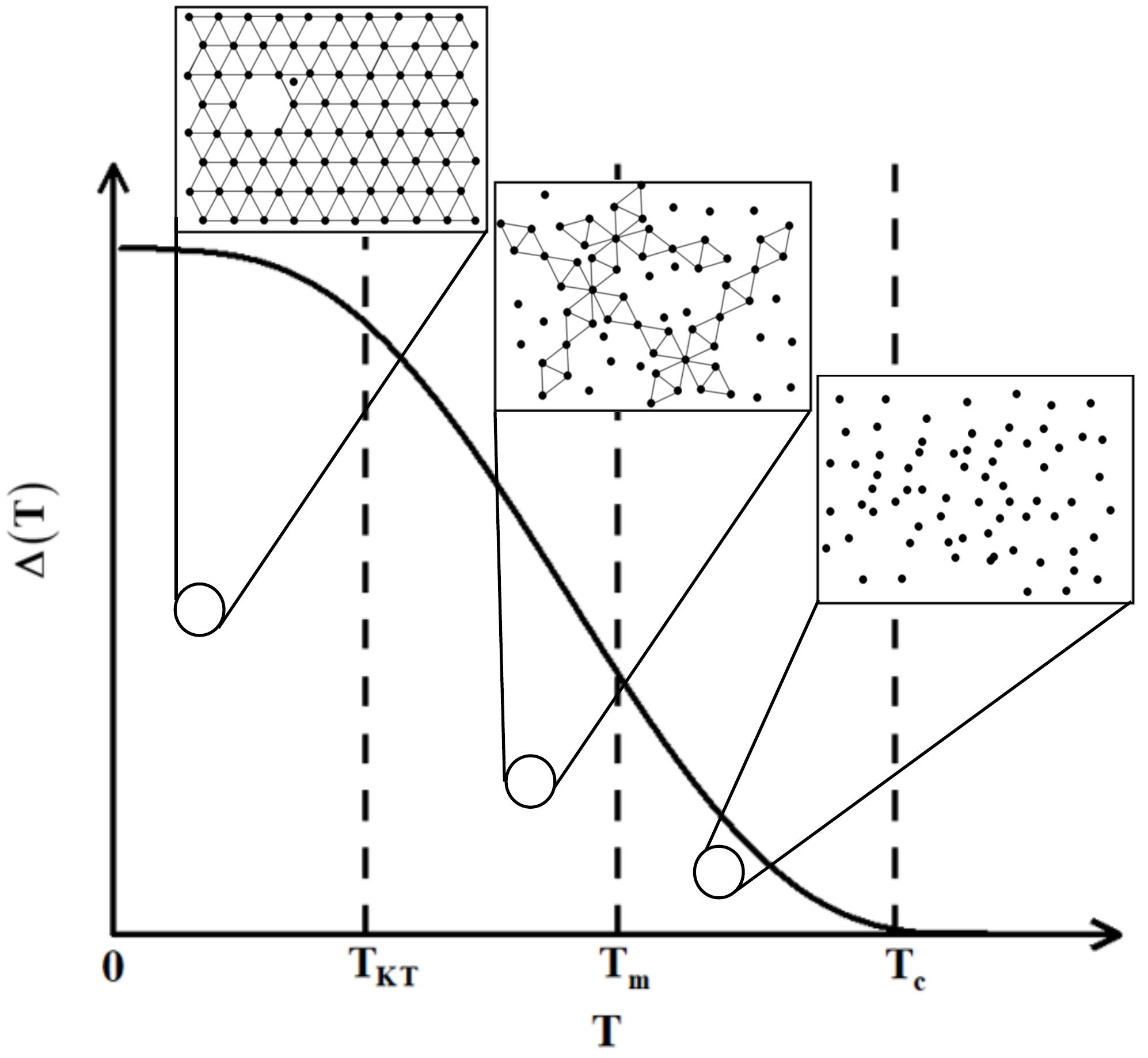}}
\protect\caption{Schematic phase diagram of the superconducting gap vs. temperature. The solid curve represents the relationship between the order parameter and temperature of the type-II superconductor under a constant external magnetic field. When $T_{KT}<T<T_m$, the vortex lattice possesses quasi-long order. Three insets show the states of the vortex lattice at diverse temperature intervals.
 }
 \label{Fig5}
\end{figure}

According to the generic nature of phase transitions in two-dimensional materials \cite{gammel1988evidence,safar1993experimental,gammel1988flux}, we can reasonably deduct the phase diagram as shown in Fig.~\ref{Fig5}. In the illustration, $T_{KT}$ is the lowest with a second melting temperature $T_m$ for the solid-liquid phase transition below the superconducting transition temperature $T_c$. For $T_{KT}<T<T_{m}$, there should exist local fragments of vortex lattice, and in each of them long-range order out to be preserved. This sets the prerequisites for the existence of vacancies and the applicability of the vacancy model proposed by Ao et.~al. \cite{ao1998motion}.

\section{Hall anomaly by vacancies of the vortex lattice}

In this section, we intend to quantitatively demonstrate that the Hall anomaly is in full concert with the theoretical basis of the vacancy hypothesis. Namely, it is due to the movement of vacancies, a direct result of many-body vortex interaction and the origin of the Hall anomaly. First, the dynamics of vacancies in a vortex lattice are discussed in detail. Then some results are applied to fit the experimental data by the Harvard group \cite{zhao2019sign} with no adjustable parameters. The vacancy activation/formation energy under a diverse set of magnetic fields is found to be the same order of magnitude as theoretical predictions. In particular, both theoretical and experimental results show that the energy of an independent vortex line or a vortex-anti-vortex pair is an order of magnitude larger than the vacancy energy.

\subsection{Properties of vacancies in a pinned vortex lattice}

In 1993, Thouless and Ao \cite{ao1993berry} proved that the existence of Magnus force is a universal essence of superconductor vortex line. In 1998, Ao \cite{ao1998motion} further established a set of processes leading to the Hall effect as a result of moving vacancies in a background of pinned vortex lattice(s). It manifests in particular that neither a modification on the vortex equation nor an assumption of two types of carriers is necessary. One only needs to study the vortex dynamics equation proposed by Niu, Ao and Thouless \cite{niu1994feynman} in 1994. To recite the theory, we turn to a crucial quantitative result regarding the motion of vacancies in a pinned vortex lattice used in the subsequent analysis, namely the vacancy formation energy in a flux line lattice.

First we look at the energy scale of dislocations in the lattice. In a type-II superconductor with mixed states, the many-body correlation between the vortices and the pinning forces usually cannot be ignored. On a two-dimensional flux line lattice (FLL) of a thin film of thickness $d$, spontaneous nucleation of a pair of edge dislocations costs an energy \cite{brandt1995flux,blatter1994vortices}
\begin{equation}\label{dislocations energy}
 e_d(r) \cong da^2\left(\frac{c_{66}}{4\pi}\right)\ln\left(\frac{r}{a}\right),
\end{equation}
where $a=(2\Phi_0/\sqrt{3}B)$ is the  lattice spacing, $r \gg a$ is the distance between two dislocations, and $c_{66}$ is the shear modulus of the FLL, cf. \cite{kosterlitz2016kosterlitz}. For uniform distortions the elastic moduli \cite{brandt1995flux} of a triangular FLL reads
\begin{equation}\label{elastic moduli}
 c_{66}\approx \left( \frac{B\phi_0}{16\pi \lambda^2\mu_0}\right)\left(1-\frac{1}{2\kappa^2}\right)\left(1-b^2\right)\left(1-0.58b+0.29b^2\right).
\end{equation}
Here $\lambda^2=(m^\ast c^2/8\pi\rho_se^2)$ is the London penetration depth (with the effective mass $m^{\ast}$ and superfluid density $\rho_s$ of the underlying carriers of charge $2e$), $\Phi_0=(hc/2\left|e\right|)$ is the flux quantum of a Cooper pair, $\kappa$ is the GL parameter and $b=H/H_{c2}$ (with the applied magnetic field strength $H$ the upper critical field $H_{c2}$ of the superconductor). The 2-d FLL is then a uniaxial elastic medium similar to that of an isotropic and bulk superconductor. In the limit of large $\kappa$ and relatively small magnetic field, we have
\begin{equation}\label{elastic moduli Re}
 c_{66}\approx \left( \frac{B\Phi_0}{16\pi \lambda^2\mu_0}\right).
\end{equation}
Using Eqs.~(\ref{dislocations energy})-(\ref{elastic moduli Re}) we can re-express the energy for a dislocation pair as
\begin{equation}\label{a dislocation pair}
 e_d(r) = \frac{d\,\varepsilon_0}{2\sqrt{3}\pi}\ln\left(\frac{r}{a}\right).
\end{equation}
Here the major variable $\varepsilon_0\equiv \left( \Phi_0/4\pi\lambda\right)^2$ defines vortex creation energy per unit-length, $(d\,\varepsilon_{0})$ then sets the scale for both the vortex-vortex and strong pinning interactions\cite{ao1998motion}. The energy scale $(\varepsilon_0/2\sqrt{3}\pi)$ for the dislocation pair is about ten times smaller than $\varepsilon_0$, it is energetically favourable to have close-distance dislocation pairs in the lattice as carriers of transverse current.

Thus at temperature $k_{B}T\ll (d\,\varepsilon_0)$ we can ignore the contribution to the current from the vortices hopping out of pinning as well as thermal activation of vortex-antivortex pairs. This is because the entire vortex lattice, formed via inter-vortex interactions should be effectively pinned down. Instead, we should look into the vacancies and interstitials which can be viewed as the smallest dislocation pairs \cite{friedel2013dislocations}. The vacancy formation energy $\varepsilon_v$ per unit length can be estimated \cite{ao1998motion} by setting $r\sim 2a$ in $e_d(r)$ together with an extra factor
\begin{equation}\label{vacancy}
 \varepsilon_v \sim \frac{\varepsilon_{0}\ln 2 }{2\sqrt{3}\pi}\left(\frac{a}{\xi}\right).
\end{equation}
Note that in Ao's initial valuations, the effect of the magnetic field on the energy barrier in the activation process has been ignored. It is anticipated that such an effect should relate to the ratio of the lattice constant as a function of a magnetic field to the coherence length $\xi$ in some way. In this work, we attempt to take the effect into account with the simplest multiplication factor $(a/\xi)$.

\subsection{Vacancy activation energy under magnetic field}

We can now analyze the experimental data from the SM of \cite{zhao2019sign} with the Arrhenius empirical formula whose validity in solid-state kinetics has been illustrated in e.g. \cite{mossel2000use}. Taking that the dominant source of resistance $R$ is the thermal activation of the dissipative vacancies in the superconductor film, we have
\begin{multline}\label{logarithm}
 \;\;R=A\exp\left(E_a/k_{B}T\right) \;\Rightarrow\; \\
 \log_{10} R(T)=\log_{10} A-(\log_{10}e)[E_a(T)/k_BT],\;\;\;\;
\end{multline}
where $A$ is a pre-factor for the exponential term, and $T$ is the temperature with $k_{B}$ the usual Boltzmann constant. Both $A$ and the activation energy $E_a$ can be themselves temperature-dependent.
Now in the GL analysis, the penetration depth $\lambda(T)$ is given by
\begin{equation}\label{lambda}
 \lambda(T)=\lambda(0)/[1-(T/T_c)]^{1/2}.
\end{equation}
The length scale is greatly affected by temperature, while its dependence on the magnetic field is negligible \cite{pippard1950field}. Substituting $\lambda(T)$ into the vacancy formation energy Eq.~(\ref{vacancy}), we get
\begin{eqnarray}
 \label{energy}
 E_a(T) &=& \frac{d\ln 2}{2\sqrt{3}\pi}\left(\frac{a}{\xi}\right)\left(\frac{\Phi_0}{4\pi\lambda(T)}\right)^2.
\end{eqnarray}

Observe that the fitting requires the magnetic field $B$ to be in the middle of the lower critical magnetic field $H_{c1}$ and the upper critical magnetic field $H_{c2}$. Although for thicker films the thickness in Eq.~(\ref{logarithm}) depends on various parameters such as magnetic field, pinning, temperature and anisotropy. The case here is relatively simple. We take $d$ to be $50\%$ of the {\em physical} thickness of the thin film, which is based on the observation that the ratio roughly corresponds to the ``superconductive'' portion of the material (along the $c$-axis).

\begin{figure}
{
\includegraphics[width=1\textwidth]{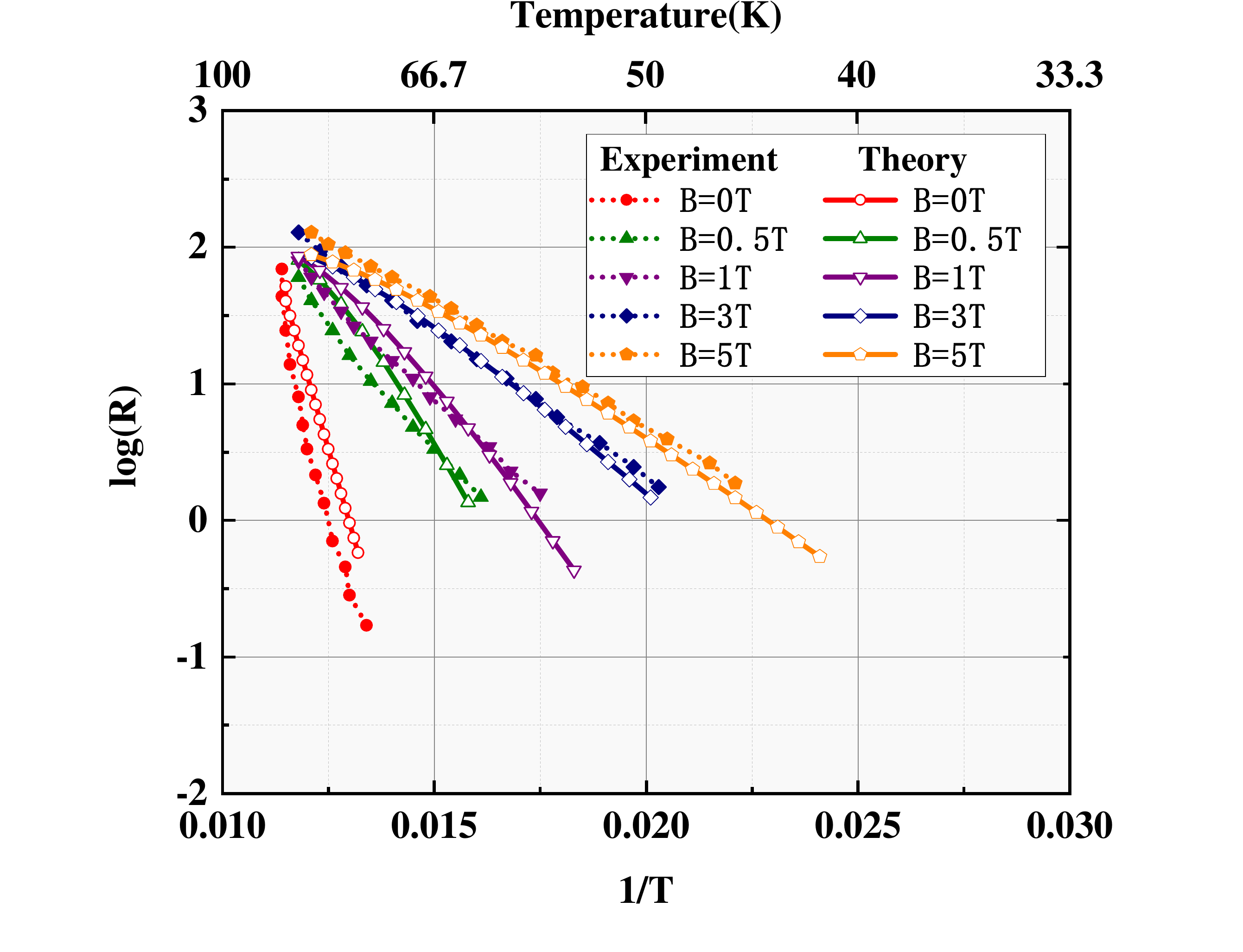}}
\protect\caption{An Arrhenius plot of the resistance vs. temperature with the solid lines and symbols from the experimental data \cite{zhao2019sign} and the open symbols and dash lines from Eq.~(\ref{logarithm}). In these dashed lines, there is an external magnetic field, the activation energy is governed by the vacancy motion, which is determined by Eq.~(\ref{energy}). In the absence of an external magnetic field, the activation energy is caused by the independent vortex motion, which depends on Eq.~(\ref{single vortex2}).
 }
 \label{Fig4}
\end{figure}

To proceed further, we set $\lambda(0)=2690$ \AA \cite{prozorov2000measurements} for BSCCO-2212, which can be compared to other values from i) reversible
magnetization, $\lambda\approx 2100$ \AA \cite{kogan1993role}; ii) uSR, $\lambda\approx 1800$ \AA \cite{lee1993evidence}; and iii) lower critical field measurements, $\lambda\approx 2700$ \AA \cite{niderost1998lower}. The superconducting transition temperature $T_c=89$ K and $\log_{10} A\approx 2$ are read off from the experimental figure \cite{zhao2019sign} for the BSCCO-2212 film. Other experimental parameters include the effective film thickness $ = 50\%$ of a 3-layer film $d=1.5UC=2\times 1.5\times 15.35\times 10^{-8}$ cm (The half height a unit cell in Bi-2212 is 15.35 \r A \cite{dou2018experimental}), the GL parameter $\kappa=86$ \cite{stintzing1997ginzburg}, the coherence length $\xi=\lambda(T)/\kappa$, together with the flux quantum of Cooper pair $\Phi_0=hc/2\left|e\right|=2.07\times 10^{-7}$ G$\cdot$ cm$^2$ and the Boltzmann constant $k_{B}=1.38\times 10^{-16}$ erg/K. All quantities are in CGS units. We then apply Eq.~(\ref{logarithm}) to Fig 7 of SM in \cite{zhao2019sign}. The result that requires no extra fitting parameter is shown in Fig. \ref{Fig4}. The theoretical values of the average energy of vacancy formation under diverse magnetic fields are presented in Table~\ref{Tab2}. Both display excellent agreement between theory and experiment, with no adjustable parameters.

\subsection{Vortex-anti-vortex pairs in the absence of a magnetic field}

In the zero-field case, the activation $E_a(T)$ is in the order of an independent vortex energy $\varepsilon_{in}$, which reads \cite{blatter1994vortices}
\begin{equation}\label{single vortex}
 \varepsilon_{in} = d\left(\frac{\Phi_0}{4\pi\lambda(T)}\right)^2\ln \kappa.
\end{equation}
For a vortex-anti-vortex pair, the distant current contribution cancels out, hence the item $\ln \kappa$ should drop out. Therefore, for an impact pair, the creation energy can be taken as
\begin{equation}\label{single vortex2}
 E_a(T) = 2d\,\varepsilon_0.
\end{equation}
Substituting the above formula along with parameters taken for the other estimates to Eq.~(\ref{logarithm}), we can compare the result to Fig 7 of SM in  \cite{zhao2019sign}. The outcome with no further variable again is shown in Fig. \ref{Fig4}, with the calculated average formation energy of vortex-anti-vortex pair in Table~\ref{Tab2}. Both the field-dependent and zero-field predictions are in excellent agreement with experimental measurements. They are further discussed below.

\begin{table}[h!]
\centering
\begin{tabular}{|c|c|c|c|c|c|}
  \hline
  B/T
  & 0  & 0.5 & 1 & 3 & 5 \\
  \hline
  E/K(Experimental value)
  & 1273 & 368.3 & 293.0 & 218.5 & 188.7 \\
  \hline
  E/K(Theoretical value)
  & 2414 & 412.5 & 330.8 & 205.6 & 172.9 \\
  \hline
 \end{tabular}
\protect\caption{The energy of the 3UC BSCCO film on various magnetic fields. The experimental value is the activation energy of the film, which is gotten by Eq.~(\ref{logarithm}). The theoretical values of the film at $B=0$ T and $B\neq 0$ T are determined by Eq.~(\ref{single vortex2}) and Eq.~(\ref{energy}), respectively.}
\label{Tab2}
\end{table}

\subsection{A comparison of vacancy and vortex-pair activation energies}

Eq.~(\ref{logarithm}) states that the slopes of the curves in Fig 7 of SM in \cite{zhao2019sign} match to $E_a/k_B$. Their average values are tabulated in Table~\ref{Tab2}. Observe that the activation energy at zero field $B=0$ is an order of magnitude larger than that at $B > 0$, which represents one of the Primary outcomes of the present paper. For $B=0$ there are no vacancies and the activation energy is at the same order of magnitude as the independent vortex energy.

On the other hand, the prime contribution to the activation energy comes from vacancies for $B\neq 0$, the influence of the independent vortex is very small and can be ignored beyond the likely melting temperature $T_{m}$, cf. Fig.~\ref{Fig5}. Keeping the same temperature, the films under a magnetic field have much larger resistance than they have under a zero field, cf. Fig 7 of SM in \cite{zhao2019sign}. The corresponding activation energies at $B\neq 0$ T are all in the same order of magnitude as the vacancy formation energies in Table 1.

By quantitative comparisons between their experimental and the theoretical values, we conclude that the Hall anomaly is a consequence of many-body vortex interactions, which further conforms to the theoretical analysis of the vacancy model. In particular, according to the above analysis and Table~\ref{Tab2}, the vortex-anti-vortex energy is an order of magnitude larger than the vacancy energy. The latter constitutes one of our primary arguments.

\section{Discussions}

The current microscopic understandings of the Hall anomaly in the 3UC BSCCO crystal may be classified into two opposite physical models, which are based on disparate theoretical approaches and give contradictory interpretations. The Harvard group \cite{ao1998motion} fabricated a few unit-cell (UC) thick BSCCO and detected the reversal depicted by vortex dynamics in this system. It laid the foundation to differentiate diverse models experimentally. The group \cite{zhao2019sign} has made a theoretical explanation based on the individual vortex dynamic model \cite{feigel1995sign}. Inversely,
Ao proposed a multi-body correlation model \cite{ao1998motion} in which vacancy movements on pinned vortex lattice are the primary mechanism for the phenomenon.

In this paper we elaborate, via quantitative analysis, on the three main aspects of the Ao's model. Firstly, the precondition and theoretical basis of the Hall anomaly warrant the vacancy model in a pinned vortex lattice or fragments of the lattice. Secondly, both theory and experiment reveal that the energy of the vortex-anti-vortex pair is about an order of magnitude higher than the vacancy energy. Last but not least, distinct theoretical models of the Hall anomaly can be quantitatively distinguished via experiment. Particularly, our theoretical predictions have no adjustable parameters. As discussed in Part E of SM \cite{supplementary}, several predictions of the vacancy model explain the Hall anomaly of BSCCO in the mixed state better than some other models.

We also look forward to providing a macro-theoretical framework for related topics in future works. For example, Yang et al. \cite{yang2022signatures} observed linear-in-temperature and linear-in-magnetic field resistance on nano-patterned YBCO film arrays over extended temperature and magnetic field. Meanwhile, the low-field magneto-resistance oscillates with a period dictated by the superconducting flux quantum.  It is possible that the unexpected signatures may be explained by a pinned vortex lattice model in a similar consideration. In particular, the plasticity of a vortex lattice pinned by the periodic nano holes can lead to diminishing of super concurrent as a result of free-energy minimisation. Further exploration will be presented elsewhere.

\begin{acknowledgments}
One of us (NG) is grateful to Prof. Chuanbing Cai for useful discussions on the research. This work was supported in part by the National Natural Science Foundation of China No. 16Z103060007 (PA).
\end{acknowledgments}

\end{document}


\title{Supplementary Material for \protect\\ Hall anomaly by vacancies in pinned lattice of vortices: A quantitative analysis on the thin-film data of BSCCO}
\author{Ruonan Guo}
\affiliation{Shanghai Center for Quantitative Life Sciences \& Physics Department, Shanghai University, Shanghai 200444, China}
\affiliation{Shanghai Key Laboratory of High Temperature Superconductors, Shanghai University, Shanghai 200444, China}
\author{Yong-Cong Chen}
\email{chenyongcong@shu.edu.cn}
\affiliation{Shanghai Center for Quantitative Life Sciences \& Physics Department, Shanghai University, Shanghai 200444, China}
\author{Ping Ao}
\email{aoping@sjtu.edu.cn}
\affiliation{Shanghai Center for Quantitative Life Sciences \& Physics Department, Shanghai University, Shanghai 200444, China}

\begin{abstract}
 This Supplementary Material (SM) covers additional computational, comparison works used or referred in the main article published in \cite{mainwork}.
\end{abstract}
\maketitle

\renewcommand{\vec}[1]{\mathbf{#1}}
\newcommand{\bvec}[1]{\mbox{\boldmath $#1$}}

\renewcommand{\theequation}{S\arabic{equation}}
\renewcommand{\thefigure}{S\arabic{figure}}
\renewcommand{\thesection}{\Alph{section}}
\renewcommand{\thesubsection}{\Alph{section}.\arabic{subsection}}
\newcommand{\cmmnt}[1]{\ignorespaces}

\renewcommand{\theequation}{S\arabic{equation}}
\renewcommand{\thefigure}{S\arabic{figure}}
\renewcommand{\thetable}{S\arabic{table}}

\tableofcontents

\section{The resistivity solutions in a phenomenological model}

The phenomenological model, which attempts to explain the Hall anomaly by adding an adjustable parameter $\alpha$ in front of the Magnus force, was first proposed by Hall and Vinen \cite{hall1956rotationII,hall1956rotation} in 1956 and later developed by Nozieres and Vinen \cite{nozieres1966motion}, Hagen et al. \cite{hagen1990anomalous,hagen1993anomalous}. This model makes a slight alteration on the vortex dynamics equation proposed by Onsager \cite{onsager1949motion} in 1949, as in Eq.~(\ref{Langevin,2}). To describe its theoretical basis, one has to derive both longitudinal $\rho_{xx}$ and Hall $\rho_{xy}$ resistivities at zero degrees and low temperatures ($T<Tc$), starting from Eq.~(\ref{Langevin,2}). The extreme scenario of a strong pining force $\vec{F}_p$ is discussed in the main article \cite{mainwork}, in this section with no pinning force $\vec{F}_p=0$.

We consider a two-fluid model of a superconductor placed at fixed temperature $T$ with a current density $\vec{J}$, the normal circuit follows a parallel connection to the superconducting circuit so that
\begin{eqnarray}
\label{22}
 \vec{J} &=& \vec{J}^S(T)+\vec{J}^N(T),\\
\label{23}
 \vec{E}(T) &=& \vec{E}^S(T) = \vec{E}^N(T).
\end{eqnarray}
Here the electric field of superfluid electrons $\vec{E}^S(T)$, normal electrons $\vec{E}^N(T)$, and the film $\vec{E}(T)$ is the same, while $\vec{J}$ is the sum of the superconducting $\vec{J}^S(T)$ and the normal $\vec{J}^N(T)$ current density. And
\begin{eqnarray}
\label{24}
 \vec{J}^S(T) &=& \sigma^S(T)\vec{E^S}(T),\\
\label{25}
 \vec{J}^N(T) &=& \sigma^N(T)\vec{E^N}(T).
\end{eqnarray}
$\sigma^N(T)$ and $\sigma^S(T)$ are the conductivity of the normal and superconducting circuit respectively, leading us to the combined conductivity
\begin{equation}\label{26}
\sigma(T) = \sigma^N(T)+\sigma^S(T).
\end{equation}
Whereas resistivity is the inverse of conductivity
\begin{eqnarray}
\label{40}
 \rho^S(T) &=& 1/\sigma^N(T),\\
\label{41}
 \rho^N(T) &=& 1/\sigma^S(T),\\
\label{42}
 \rho(T) &=& 1/\sigma(T),
\end{eqnarray}
where $\rho(T)$, $\rho^N(T)$ and $\rho^S(T)$ are, respectively, the film, normal electrons and superfluid electrons resistivity. In a true superconductor, the resistivity and conductivity are determined by a matrix, and the non-diagonal elements are antisymmetric.

\begin{figure}[!ht]
\begin{tabular}{cc}
{\label{supplement_1.pdf}
\includegraphics[width=0.5\textwidth]{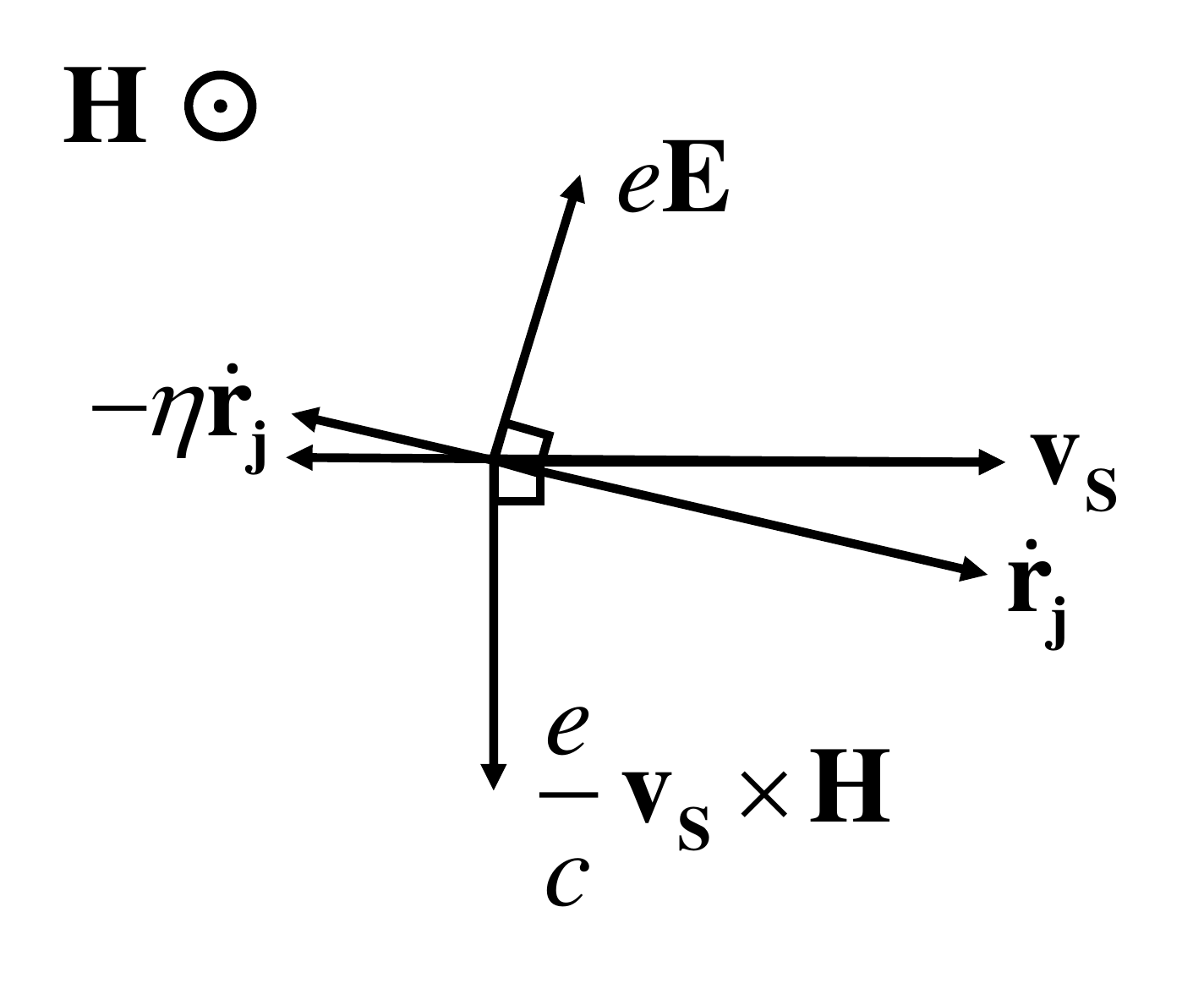}}&
{\label{supplement_2.pdf}
\includegraphics[width=0.5\textwidth]{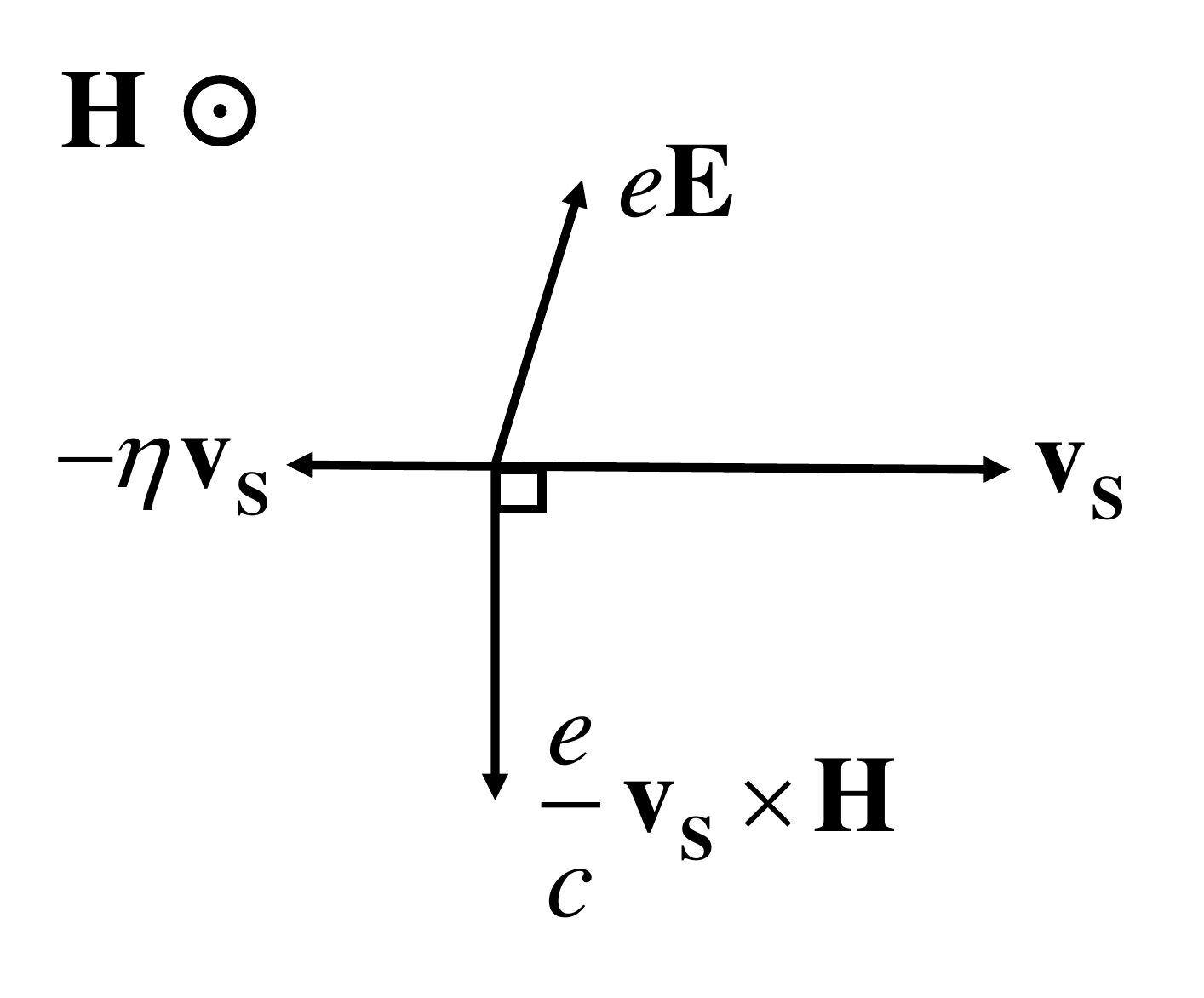}}\\
(a) & (b)
\end{tabular}
\protect\caption{With no pinning force $\vec{F}_p=0$, the dynamic analysis of vortex and normal electron at temperature $T$. In the picture $\vec{E}$ is the electric field generated by moving vortex and $\vec{H}$ is the applied magnetic field. (a) The moving vortex generates resistivity at temperature $T$ ($0\leqslant T<T_c$). (b) Partial resistivity comes from the normal component of electrons at temperature $T$ ($0<T\leqslant T_c$).
}
\label{supplement}
\end{figure}

For zero degrees, the total carriers become superfluid electrons, i.e., the normal electron density $n_n(0)=0$, hence $\sigma^N(0)=0$. Then, the film conductivity is the superconducting contribution $\sigma^S(0)$
\begin{equation}\label{27}
\sigma(0) = \sigma^S(0).
\end{equation}
In a steady-state where the vortex dynamics analysis is shown in Fig.~\ref{supplement}(a), and the dynamic equation is
\begin{equation}\label{1}
q\left(\frac{n_s(0)}{2}\right)h(\vec{v}_{s,t}(0)-\alpha\dot{\vec{r}}_j(0))\times \hat{\vec{z}}-\eta\dot{\vec{r}}_j(0) = 0.
\end{equation}
The physical significance of these symbols is detailed in the main article \cite{mainwork}. The superfluid electron moves with a velocity $\vec{v}_{s,t}(0)=\vec{J}/2en_{s}(0)$ under the applied transport current $\vec{J}$, where $n_{s}(0)$ is the density of superfluid electron at zero temperature. For calculation purposes, according to Fig.~\ref{supplement}(a), with the direction parallel and perpendicular to the superfluid electron motion as the base vector, the vector Eq.~(\ref{1}) decomposed into two scalar equations by
\begin{eqnarray}
\label{2}
 -q\frac{n_{s}(0)}{2}hv_{s,t}(0)+ \alpha q\frac{n_{s}(0)}{2}h\dot{r}_{jxx}(0)+\eta\dot{r}_{jxy}(0) &=& 0,\\
\label{3}
 \alpha q\frac{n_{s}(0)}{2}h\dot{r}_{jxy}(0)-\eta\dot{r}_{jxx}(0) &=& 0.
\end{eqnarray}
Here $\dot{r}_{jxx}(0)$ and $\dot{r}_{jxy}(0)$ denote the components of the vortex velocity parallel and perpendicular to the direction of superfluid electron, respectively. The macroscopic electric field is equal to $\vec{E}=-\dot{\vec{r}}_j\times \vec{H}/c$ \cite{josephson1965potential} (satisfies the usual right-hand rules), one finds the longitudinal $E^S_{xx}(0)$ and Hall $E^S_{xy}(0)$ electric fields
\begin{eqnarray}
\label{6}
 E^S_{xx}(0) &=& -\dot{r}_{jxy}(0)H/c,\\
\label{7}
 E^S_{xy}(0) &=& \dot{r}_{jxx}(0)H/c.
\end{eqnarray}
Thus the longitudinal $\sigma_{xx}(0)$ and Hall resistivities $\sigma_{xy}(0)$ are immediately obtained
\begin{eqnarray}
\label{43}
 \sigma_{xx}(0) &=& -\frac{ec(\alpha^2q^2n_{s}^2(0)h^2+4\eta^2)}{\eta qhH},\\
\label{44}
 \sigma_{xy}(0) &=& \frac{2ec(\alpha^2q^2n_{s}^2(0)h^2+4\eta^2)}{\alpha n_{s}(0)q^2h^2H},
\end{eqnarray}
then we have the longitudinal $\rho_{xx}(0)$ and Hall resistivities $\rho_{xy}(0)$
\begin{eqnarray}
\label{8}
 \rho_{xx}(0) &=& -\frac{\eta qhH}{ec(\alpha^2q^2n_{s}^2(0)h^2+4\eta^2)},\\
\label{9}
 \rho_{xy}(0) &=& \frac{\alpha n_{s}(0)Hq^2h^2}{2ec(\alpha^2q^2n_{s}^2(0)h^2+4\eta^2)}.
\end{eqnarray}
This allows the sign reversal of the Hall resistivity $\rho_{xy}(0)$ when $\alpha<0$.

We now proceed to discuss the low-temperature ($0<T<T_c$) scenario. There should be resistivity coming from the normal and superfluid components. In terms of Eq.~(\ref{8}) and Eq.~(\ref{9}), the superfluid contribution at $T$ is
\begin{eqnarray}
\label{45}
 \sigma_{xx}(T) &=& -\frac{ec(\alpha^2q^2n_{s}^2(T)h^2+4\eta^2)}{\eta qhH},\\
\label{46}
 \sigma_{xy}(T) &=& \frac{2ec(\alpha^2q^2n_{s}^2(T)h^2+4\eta^2)}{\alpha n_{s}(T)q^2h^2H},
\end{eqnarray}
Under Eq.~(\ref{24}) and Eq.~(\ref{25}), we get the normal current density $\vec{J}^N(T)$. The normal electron moves with an average velocity $\vec{v}_{n}=\vec{J}^N(T)/en_{n}(T)$, with the dynamic analysis is illustrated in Fig.~\ref{supplement}(b), and the steady-state equation is
\begin{equation}\label{10}
 \frac{e}{c}\vec{v}_n\times\vec{H}+e\vec{E}-\eta'\vec{v}_n = 0.
\end{equation}
The term $-\eta'\vec{v}_n$ is the frictional force on the normal electron. In line with the procedure in Eqs.~(\ref{1})-(\ref{3}), the vector Eq.~(\ref{10}) decomposed into
\begin{eqnarray}
\label{32}
 eE^N_{xy}(T) &=& ev_n(T)H/c,\\
\label{33}
 eE^N_{xx}(T) &=& \eta'v_n(T).
\end{eqnarray}
Where $E^N_{xx}(T)$ and $E^N_{xy}(T)$ present the longitudinal and Hall electric fields of the normal electrons, separately. The solutions of Eq.~(\ref{45}-\ref{33}) are the normal state longitudinal conductivity $\sigma^N_{xx}(T)$ and Hall conductivity $\sigma^N_{xy}(T)$
\begin{eqnarray}
\label{13}
 \sigma^N_{xx}(T) &=& e^2n_{n}(T)/\eta',\\
\label{14}
 \sigma^N_{xy}(T) &=& ecn_{n}(T)/H.
\end{eqnarray}
Here the density of normal electron $n_n(T)$ can be written as $n_s(0)-n_s(T)$. Since the superconducting circuit is connected in parallel with the normal circuit, the longitudinal $\sigma_{xx}(T)$ and Hall $\sigma_{xy}(T)$ conductivity of the film read
\begin{eqnarray}
\label{15}
 \sigma_{xx}(T) &=& -\frac{e\left[c\eta'\alpha^2q^2n_s^2(T)h^2-eh\eta qH(n_s(0)-n_s(T))+4c\eta'\eta^2\right]}{\eta\eta'qhH},\\
\label{16}
 \sigma_{xy}(T) &=& \frac{ce\left[2\alpha^2q^2n_s^2(T)h^2+\alpha n_s(T)q^2h^2(n_s(0)-n_s(T))+8\eta^2\right]}{\alpha Hn_s(T)q^2h^2},
\end{eqnarray}
therefore
\begin{eqnarray}
\label{15}
 \rho_{xx}(T) &=& -\frac{\eta\eta'qhH}{e\left[c\eta'\alpha^2q^2n_s^2(T)h^2-eh\eta qH(n_s(0)-n_s(T))+4c\eta'\eta^2\right]},\\
\label{16}
 \rho_{xy}(T) &=& \frac{\alpha Hn_s(T)q^2h^2}{ce\left[2\alpha^2q^2n_s^2(T)h^2+\alpha n_s(T)q^2h^2(n_s(0)-n_s(T))+8\eta^2\right]}.
\end{eqnarray}
The change in the sign of $\alpha$ has to overcome the normal electron contribution to achieve the sign reversal on Hall resistivity $\rho_{xy}(T)$.

\section{A power-law behavior with $v=2$}

As pointed out in the main article \cite{mainwork}, our analysis elucidates the theoretical basis and prerequisites of the vacancy model. Now we examine its prediction. In the low-temperature limit, both longitudinal and Hall resistivities vanish exponentially \cite{ao1998motion}
\begin{equation}\label{17}
 \rho_{xy} \propto \rho^v_{xx}.
\end{equation}
Introduce the Hall conductivity,
\begin{equation}\label{18}
 \sigma_{xy} = \frac{\rho_{xy}}{\rho^2_{xx}+\rho^2_{xy}}.
\end{equation}
Whereas $\rho_{xy}\ll \rho_{xx}\ll \rho_{0}$ in real superconductors, $\rho_{0}$ is the resistivity at room temperature. The Hall conductivity term (\ref{18}) becomes
\begin{equation}\label{19}
 \sigma_{xy} = \rho_{xy}/\rho^2_{xx}.
\end{equation}
The scaling relation reads after some manipulations of Eq.~(\ref{17}) and Eq.~(\ref{19}),
\begin{equation}\label{20}
 \sigma_{xy} \propto \rho^{-(2-v)}_{xx}.
\end{equation}

The model predicts the power
\begin{equation}\label{21}
 v = \frac{2a_v+b_v}{a_v+b_v}
\end{equation}
varying between 1 and 2, depending on the detail of a sample which determines the numerical factors $a_v$ and $b_v$. If all vacancies produced by pinnings, the $b_v=0$ and $v=2$. Consider that the longitudinal resistivity obeys $\rho_{xx}(T)=\rho_0 e^{-E_a/kT}$, where $\rho_0$ is a constant. Thus, the Hall conductivity is independent of temperature when $v=2$, in agreement with Fig. 1 of \cite{zhao2020zhao}. In the illustration, the Hall conductivity is temperature independent at the low temperature, especially at a fixed magnetic field $B=6T$. On BSCCO films, the work of Zhao el al. \cite{zhao2020zhao} revealed such a power-law behavior with $v=2$. Data from other laboratories \cite{shi1,shi2,shi3,shi4,shi5} also admit $1<v<2$.

%